\documentclass[12pt]{iopart}

\usepackage{iopams}
\usepackage{graphicx}

\begin{document}

\title{Semi-classical and quantum Rabi models: in celebration of 80 years}

\author{Daniel Braak$^1$, Qing-Hu Chen$^2$, Murray T. Batchelor$^{3,4}$ and Enrique Solano$^{5,6}$}

\address{$^1$ EP VI and Center for Electronic Correlations and Magnetism, University of Augsburg, 86135 Augsburg, Germany}

\address{$^2$ Department of Physics, Zhejiang University, Hangzhou 310027, China and Collaborative Innovation Center of Advanced Microstructures, Nanjing 210093, China}

\address{$^3$ Centre for Modern Physics, Chongqing University, Chongqing 400044, China} 

\address{$^4$ Mathematical Sciences Institute and Department of Theoretical Physics, Research School of Physics and Engineering, Australian National University, Canberra ACT 2601, Australia} 

\address{$^5$ Department of Physical Chemistry, University of the Basque Country UPV/EHU, Apartado 644, 48080 Bilbao, Spain}

\address{$^6$ IKERBASQUE, Basque Foundation for Science, Maria Diaz de Haro 3, 48013 Bilbao, Spain}

\begin{abstract}
This is an introduction to the special issue collection of articles on 
{\em Semi-classical and quantum Rabi models} to be published in J.~Phys.~A to mark the 80th anniversary of the Rabi model.
\end{abstract}

In the springtime of quantum mechanics, in 1936, I.I. Rabi investigated the semi-classical version of a model which now bears his name~\cite{Rabi36}. The Rabi model describes the simplest interaction between a two-level atom and a classical light field. The fully quantized version was considered by E.T. Jaynes and F.W. Cummings in 1963 for the case where the rotating-wave approximation can be applied~\cite{JC63}. 

The quantum Rabi model has Hamiltonian 

\begin{equation}
\label{HRabi}
H_{\mathrm{QRM}} = \case12 \omega_0^R \sigma_z + \omega^R a^\dagger a + g \sigma_x (a+a^\dagger).
\end{equation}
Here $\sigma_z$ and $\sigma_x$ are Pauli matrices with $a^\dagger$ and $a$ the creation and destruction operators for the single bosonic mode. The light-matter interaction is controlled by the coupling parameter $g$, with $\omega^R$ the mode frequency and $\omega_0^R$ the qubit frequency. Distinct dynamics for different regimes are defined by the relation among these three parameters (see figure 1).

The restricted model, known as the Jaynes-Cummings model, has always shadowed the more general aspects of the quantum Rabi model. This is because the conditions for applying the rotating-wave approximation to the quantum Rabi model, which leads to the Jaynes-Cummings model, are directly relevant to most experimental regimes, where $g/\omega^R$ is small.

Moreover, the Jaynes-Cummings model is easy to solve, allowing the physical properties of the model to be readily obtained and compared with a large number of experiments. The basic physics explored in the Jaynes-Cummings model alone has many remarkable facets, including Rabi oscillations, collapses and revivals of quantum state populations, quadrature squeezing, entanglement, Schr\"odinger cat states and photon anti-bunching~\cite{Nobel2012}.

\begin{figure}[t]
\begin{center}
\includegraphics[width=0.7\columnwidth]{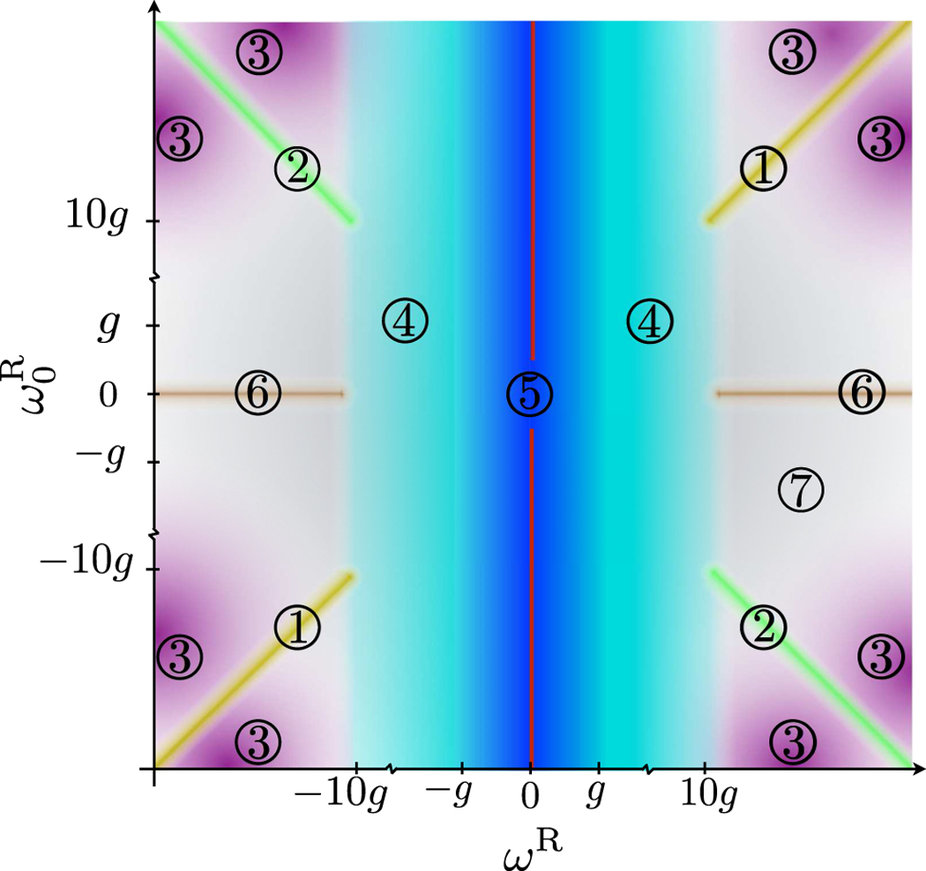}
\caption{The different coupling regimes in the configuration space of frequencies associated with the quantum Rabi model of Eq.~(\ref{HRabi}). Regions 4 and 5 define the ultrastrong and deep strong coupling regimes, respectively, in which the Jaynes-Cummings model is not applicable. 
Reproduced with permission from \cite{Pedernales}.} 
\label{fig1}
\end{center}
\end{figure}

However, the situation has recently changed with the quantum Rabi model stepping into the spotlight. 
There are three basic reasons: (i) experiments have now been able to push into the ultrastrong 
($|\omega^R| < 10g$)~\cite{Gunter,Niemczyk-Forn-Diaz} and deep strong ($|\omega^R| < g$) \cite{Yoshihara} 
coupling regimes in which the rotating-wave approximation and thus the Jaynes-Cummings model are not applicable; 
(ii) there are exciting prospects for novel regimes of light-matter interactions~\cite{Casanova} and potential applications 
in quantum information technologies~\cite{Romero}; 
(iii) an analytic solution was obtained for the full quantum Rabi model \cite{Braak, Chen, Zhong}, followed by further progress on solving various extensions of it. It is noteworthy to mention that the conceptual and experimental advances call for deeper mathematical analysis, with no restrictions on the possible values of any model parameter.

Circuit quantum electrodynamics, two-dimensional electron gases, and trapped ions, supported by the concepts of quantum metamaterials and quantum simulations \cite{Pedernales}, have emerged as platforms for faithful representations of abstract models, quite analogous to the possibilities established by the advent of cold atom physics. The impact on practical applications is even greater, because the two-level system appearing in the semi-classical and quantum Rabi models is a qubit, the building block of quantum information technologies with the ultimate goal to realize quantum simulations and quantum computations. These Rabi models thus form the connecting link in this interplay of mathematics, physics, and technology. The novel applications of certain branches of classical mathematics to quantum physics may even inspire efforts to attack hitherto unsolved problems within pure mathematics itself \cite{Wakayama}, whereas other relevant developments, e.g., the theory of quasi-exact eigenstates \cite{Turbiner}, may have direct implications for
quantum information theory.

\begin{figure}[t]
\begin{center}
\includegraphics[width=0.8\columnwidth]{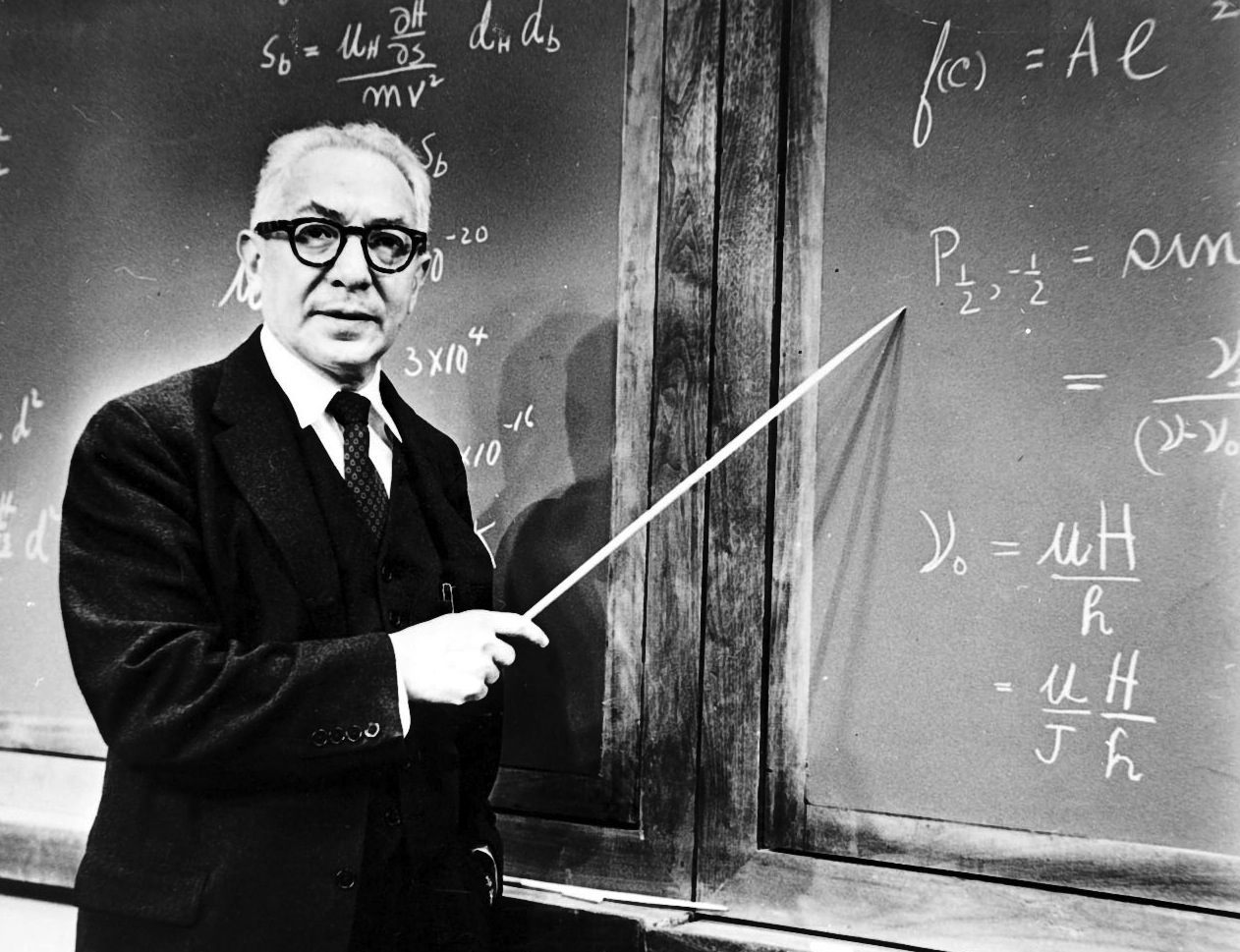}
\caption{Isidor Isaac Rabi (1898-1988) was awarded the Nobel Prize in Physics in 1944 for discovering nuclear magnetic resonance. His name is associated with a number of key concepts in physics, including the Rabi model which describes the interaction between a two-level atom and a light field. Photo credit: University Archives, Rare Book \& Manuscript Library, Columbia University in the City of New York.}
\end{center}
\end{figure}

This special collection of articles is dedicated to the semi-classical and quantum Rabi models, with all their variants and possible extensions, including nontrivial multi-qubit and multi-photon scenarios. In addition to marking the 80th anniversary of the (semi-classical) Rabi model, we anticipate that this collection of articles will
inspire further developments in this important area of cross-disciplinary research.

\end{document}